\documentclass[aps,prl,twocolumn,nopacs,superscriptaddress]{revtex4}
\usepackage{graphicx}
\usepackage{verbatim}
\usepackage{amsmath}
\usepackage{sidecap}
\usepackage{epstopdf}
\usepackage{braket}

\usepackage[usenames,dvipsnames]{xcolor}
\definecolor{Highlight}{rgb}{1,1,0.5}
\usepackage{soul}

\begin{document}


\title{Charge transfer excitations in VUV and soft X-ray resonant scattering spectroscopies}

\author{Edwin Augustin}
\author{Haowei He}
\affiliation{Department of Physics, New York University, New York, New York 10003, USA}
\author{Lin Miao}
\affiliation{Department of Physics, New York University, New York, New York 10003, USA}
\affiliation{Advanced Light Source, Lawrence Berkeley National Laboratory, Berkeley, CA 94720, USA}
\author{Yi-De Chuang}
\author{Zahid Hussain}
\affiliation{Advanced Light Source, Lawrence Berkeley National Laboratory, Berkeley, CA 94720, USA}
\author{L. Andrew Wray}
\email{lawray@nyu.edu}
\thanks{Corresponding author}
\affiliation{Department of Physics, New York University, New York, New York 10003, USA}

\begin{abstract}

The utility of resonant scattering for identifying electronic symmetries and density distributions changes dramatically as a function of photon energy. In the hard X-ray regime, strong core hole monopole potentials tend to produce X-ray absorption features with well defined electron number on the scattering site. By contrast, in the vacuum ultraviolet (VUV), resonant scattering from Mott insulators tends to reveal spectra that are characteristic of only the nominal valence, and are insensitive to deviations from nominal valence brought on by metal-ligand hybridization. Here, atomic multiplet simulations are used to investigate the interplay of monopolar and mulitpolar Coulomb interactions in the VUV and soft X-ray regimes, to identify how charge transfer thresholds and other signatures of mixed valence can manifest in this low photon energy regime. The study focuses on the Mott insulator NiO as a well characterized model system, and extrapolates interactions into non-physical regimes to identify principles that shape the spectral features.





\end{abstract}


\date{\today}

\maketitle

\section{Introduction}

\begin{figure}[t]
\includegraphics[width = 8.5cm]{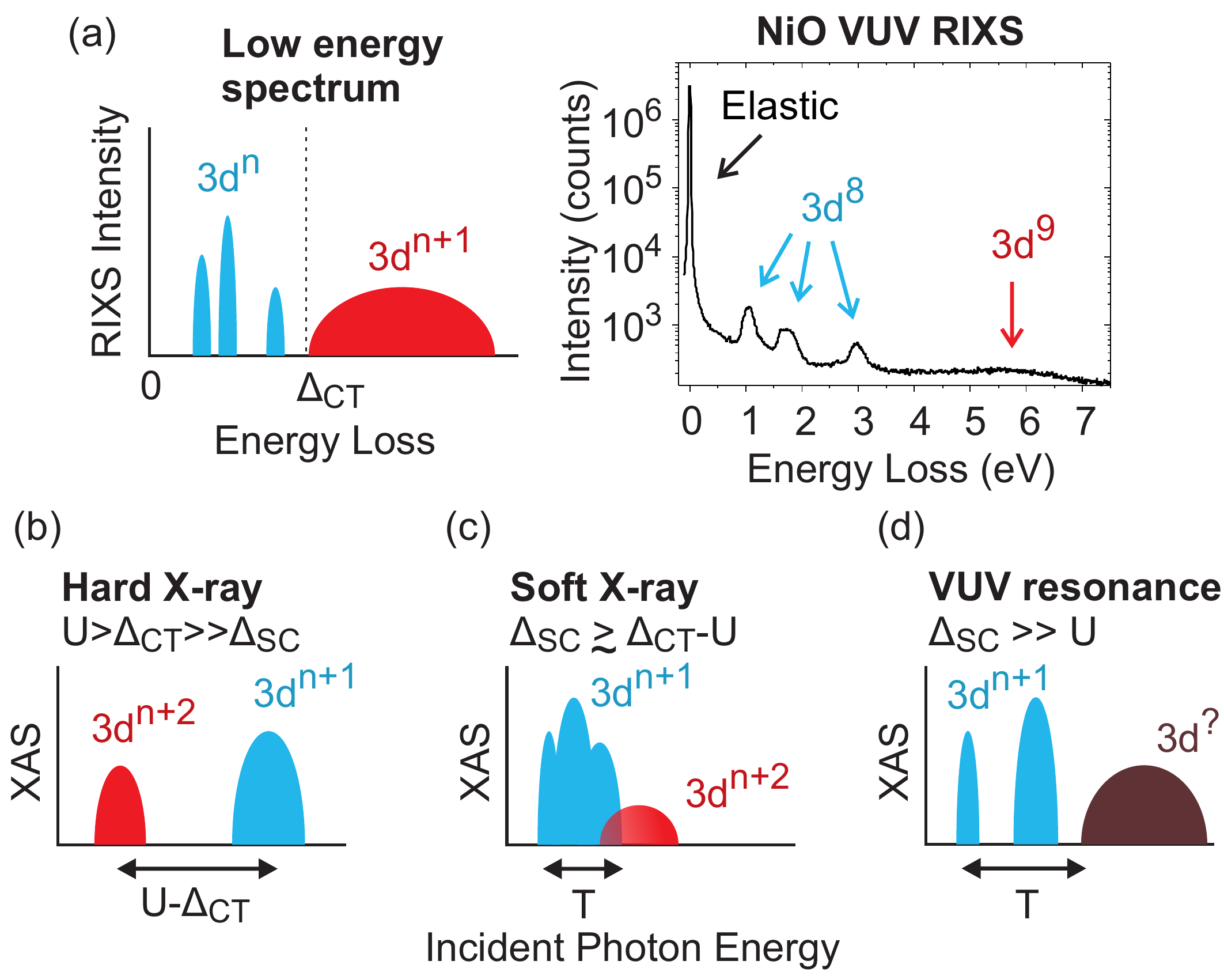}
\caption{{\bf{Energetic regimes for resonance}}: (a) (left) Energetically separated (blue) multiplet and (red) charge transfer excitations are illustrated next to (right) the nickel M-edge VUV RIXS spectrum of Mott insulating NiO, measured at $h\nu=71.5$ eV. As analyzed in Ref. \cite{WrayFrontiers}, features above the charge transfer gap ($\Delta_{CT}$) are extremely weak at the M-edge, in spite of a nominal nickel valence state of 8.2. (b-d) Schematics of X-ray absorption features are shown for the hard X-ray, soft X-ray and VUV regimes. The charge transfer threshold T is the energy gap between the first dipole allowed resonance state and the charge transfer continuum, and is given an empirical definition in the discussion of Fig. \ref{fig:LedgeTH}. In the VUV case, nominal valence multiplet features can appear above the charge transfer threshold, possibly with a mixed valence character (``$3d^?$" peak). We note that though RIXS and XAS spectra can both often be interpreted in terms of features that preserve or change nominal valence, RIXS provides spectroscopic information that cannot be obtained from XAS.}
\label{fig:cartoons}
\end{figure}

The incident energy dependence of resonant X-ray scattering tends to be determined by very different physical processes in the hard and soft X-ray regimes. Deeply bound core holes created by hard X-rays provide a very strong monopolar Coulomb perturbation to the valence level, splitting X-ray absorption (XAS) spectra into features with rather well defined valence on the scattering site \cite{KotaniDeGrootBook,kotaniAbsValence}, and often causing charge transfer states to be lower in energy than the nominal-valence resonance peaks (see Fig. \ref{fig:cartoons}(b)). In soft X-ray experiments, this monopolar core hole potential is a relatively weak perturbation, and direct wavefunction overlap between a shallow core hole and valence orbitals causes higher order angular momentum interactions to become significant (see Fig. \ref{fig:cartoons}(c)). The resulting peak structure can be analyzed to obtain detailed information about ground state symmetries and energetics, but the strong overlapping of features and loss of energetic dominance from the core hole potential make it difficult to quantitatively determine charge density on the scattering site.

This ambiguity is further accentuated for the even shallower core holes accessed in the vacuum ultraviolet (VUV). RIXS and XAS spectra of Mott insulators measured in the VUV tend to be well reproduced by renormalized nominal-valence models that completely neglect charge transfer from ligands. This is true even in cases where ligand hybridization is known to significantly increase valence electron density on the scattering atom (e.g. from 8 to 8.2 in NiO \cite{NiOp2holes}). For example, the M-edge VUV resonant inelastic X-ray scattering (RIXS) spectrum of NiO in Fig. \ref{fig:cartoons}(a, right), shows a far weaker charge transfer peak than L-edge soft X-ray RIXS \cite{WrayFrontiers,NiO_LRIXS,NiO_MRIXS,PattheyML,KotaniSIAM}, and M-edge XAS from NiO has no obvious charge transfer derived spectral feature. Here, we will explore the question of how charge transfer states manifest in the VUV by varying the core hole modeling parameters of an atomic multiplet (AM) simulation for NiO, augmented by ligand hybridization treated in the single Anderson impurity model (SAIM).

\begin{figure}[t]
\includegraphics[width = 8.7cm]{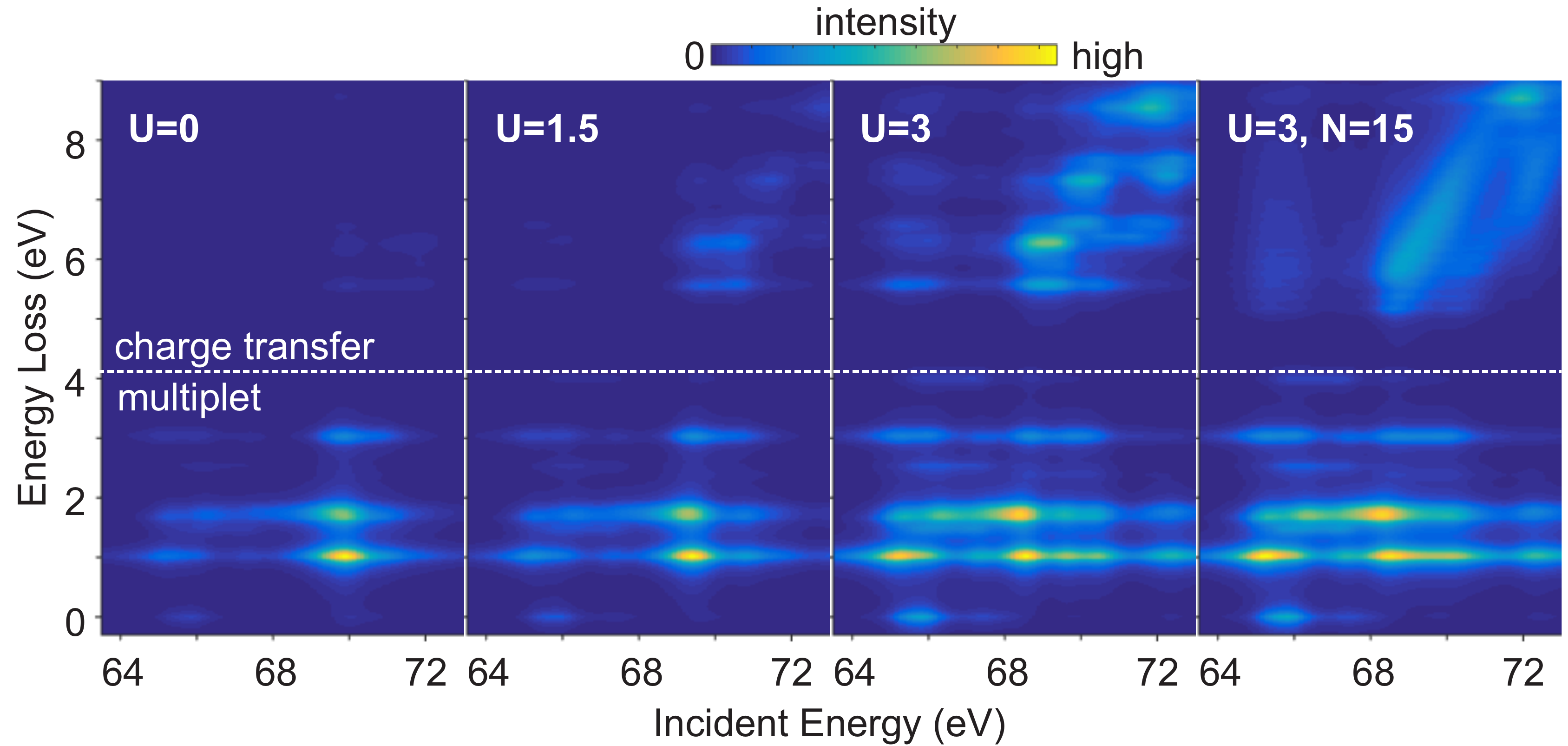}
\caption{{\bf{Simulated M-edge RIXS images}}: The RIXS spectrum of NiO is shown for different values of the core hole potential $U$. A dashed horizontal line indicates the partition between low energy multiplet excitations and higher energy charge transfer excitations. The rightmost panel uses a larger number of ligand band states (N=15) to show the well converged form of the charge transfer features.}
\label{fig:RIXSimages}
\end{figure} 

\begin{figure}[t]
\includegraphics[width = 8.5cm]{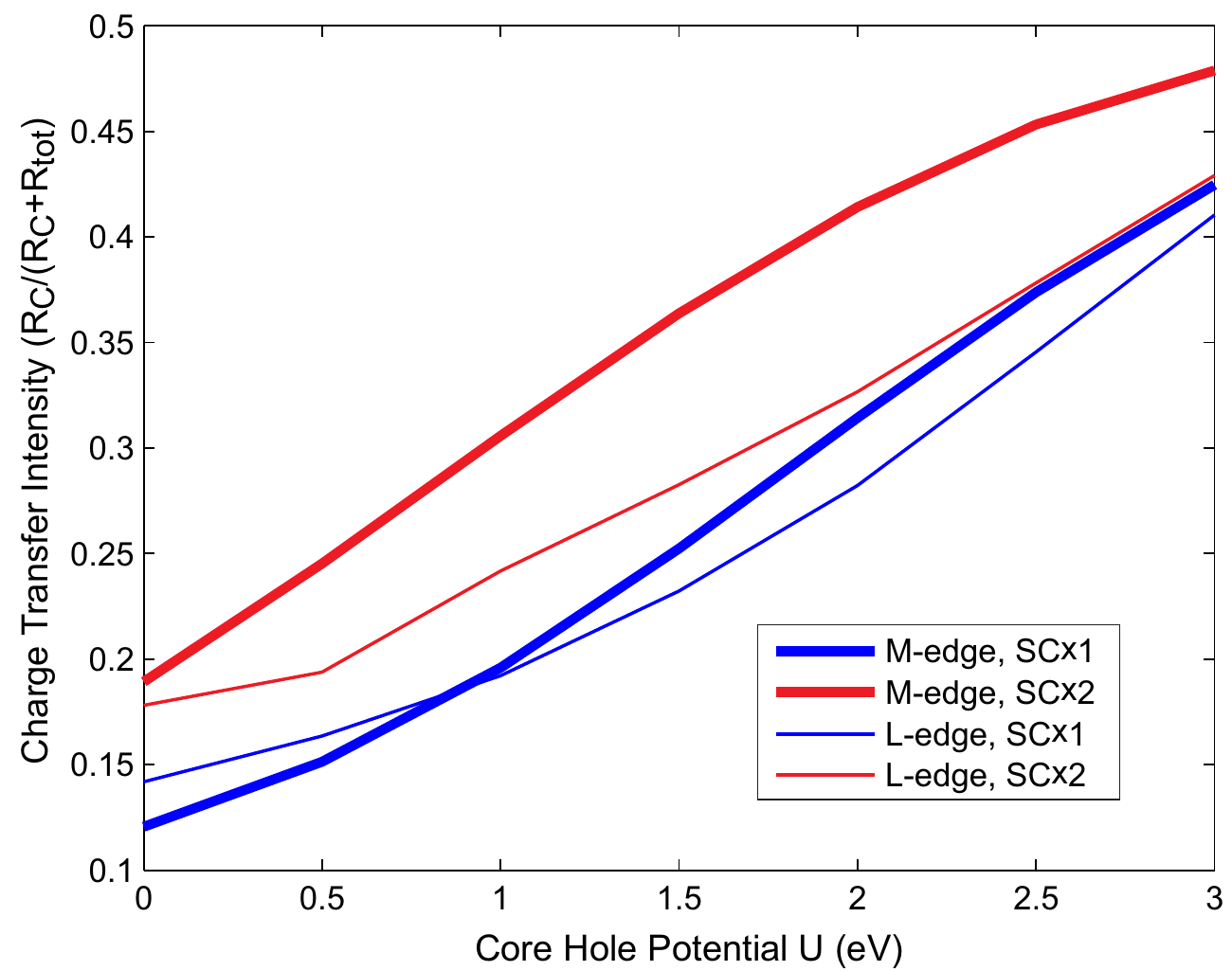}
\caption{{\bf{Charge transfer shake-up scattering intensity vs Coulomb potentials}}: The charge transfer excitation intensity (summed energy loss $E>4$ eV RIXS intensity) is plotted as a function of the core hole potential $U$. Charge transfer scattering intensity is averaged over incident energy at the M- and L- edges, and plotted as a ratio of total scattering at the relevant edge. Curves are shown for the typical Hartree-Fock derived Slater-Condon multiplet interaction strength ($SC\times1$) and for a non-physical case in which the interaction strength has been artificially doubled ($SC\times2$).}
\label{fig:shakeVsU}
\end{figure} 

\section{Methods}




The basic parameters of the AM calculation are similar to those used in Ref. \cite{KotaniSIAM,WrayNiO,WrayFrontiers}. The crystal field is described by the parameter 10Dq=0.56 eV, and the Slater-Condon parameters that govern 2-particle angular momentum coupling are set to atomic values (80$\%$ of bare Hartree-Fock values). Spin orbit coupling of core and valence Ni electrons is set to Hartree-Fock values. Core hole inverse lifetimes are set to $\Gamma=1$ eV at the M-edge, $\Gamma=0.6$ eV at L$_3$ and $\Gamma=0.8$ eV at L$_2$. Hybridization with oxygen is accounted for in he SAIM model as in Ref. \cite{KotaniSIAM}, with 0 or 1 ligand holes, a ligand band width W=3 eV, and ligand states distributed in just 3 discrete energies (N=3). The configuration energy of ligand hole states was set to 3.5 eV, in the absence of a core hole. Hopping between $e_g$ orbitals and the nearest neighbor ligand states with corresponding symmetry is $V_{eg}$=2.2 eV ($V_{t2g}=-V_{eg}/2$), and is reduced by 10$\%$ when a core hole is present. The ground state density of oxygen holes within this model matches the experimentally based estimate of 0.2 \cite{NiOp2holes}. 

The effective core hole potential is defined as $U=U_{cv}-U_{vv}$, where $U_{cv}$ is the core-valence monopole interaction (3p-3d interaction, for the NiO M-edge), and $U_{vv}$ is the same-site Mott-Hubbard interaction between valence electrons. This potential assumes larger positive values for deeper core holes, as the radial wavefunction of the core electron becomes more compact. For shallow core holes accessed in the VUV, the radial size of core level wavefunctions is quite similar to the radial size of valence orbitals, and the $U$ potential may be negligible. In the modeling implementation used for this manuscript, the valence configuration energies (average energy for a given 3d electron number) are defined entirely by $U$, the crystal field, and the ligand site energy. The dependence of spectra on `multiplet' angular momentum interactions is explored by multiplying the higher order Slater-Condon terms by a factor of 2 in some panels. This is indicated as ``SC$\times$2", and refers to the $n>0$ Slater-Condon terms (G$^n$ and F$^n$). Though the core hole potential $U$ is associated with the lowest order Slater-Condon term (F$^0$), an accurate renormalized value of the $U$ parameter can not be easily obtained from first principles \cite{MultLigWannier}, and we adopt the common practice of using the term ``Slater-Condon parameters" to reference only the higher order terms with n$>$0.

\begin{figure*}[t]
\includegraphics[width = 15cm]{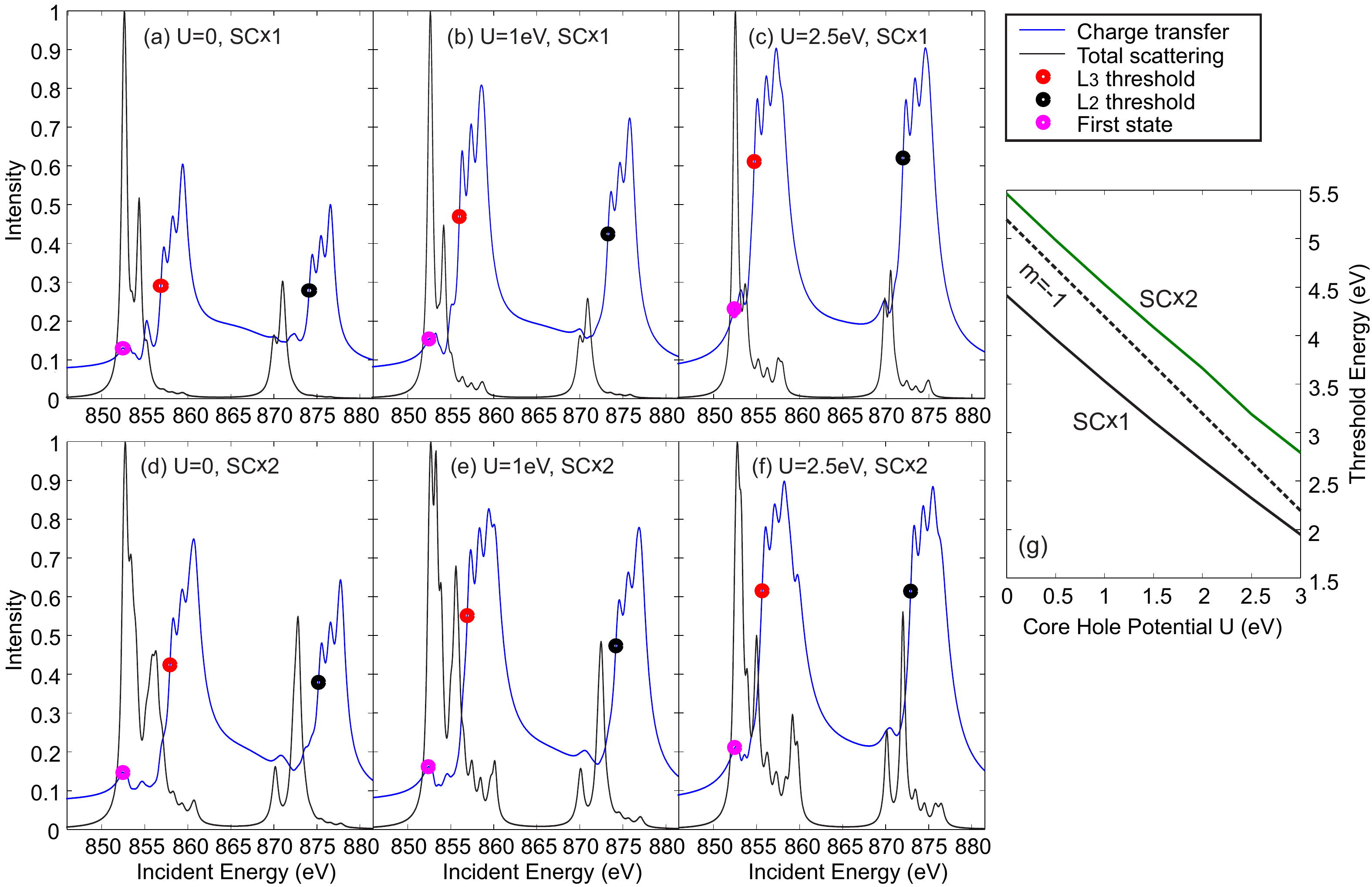}
\caption{{\bf{The charge transfer shake-up threshold in soft X-ray RIXS}}: (a-c) The L-edge XAS spectrum is plotted in black for different core hole potential $U$ values, and the fraction of RIXS intensity attributed to charge transfer (energy loss $E>4.2$ eV RIXS intensity) is plotted in blue. Circles indicate (magenta) the first resonance state energy, (red) the $L_3$ charge transfer threshold and (black) the $L_2$ charge transfer threshold, as defined in the text. Panels (d-f) show the same spectra for artificially doubled Slater-Condon values ($SC\times2$). (g) The energy difference between the first state and the $L_3$ threshold energy is plotted as a function of the core hole potential, and compared with (dashed line) a slope of -1.}
\label{fig:LedgeTH}
\end{figure*} 


\begin{figure*}[t]
\includegraphics[width = 15cm]{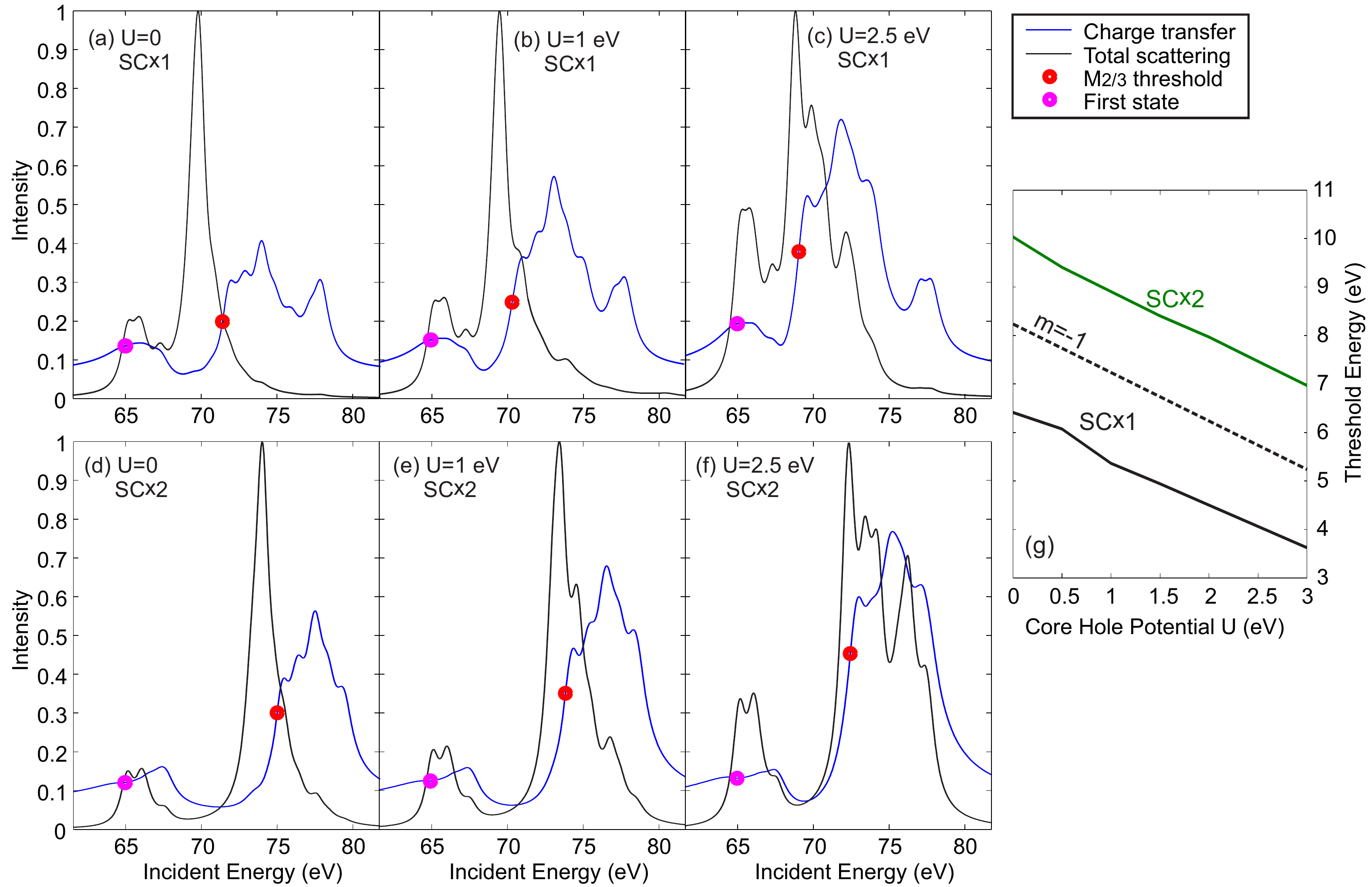}
\caption{{\bf{The charge transfer shake-up threshold in VUV RIXS}}: (a-c) The M-edge XAS spectrum is plotted in black for different core hole $U$ values, and the fraction of RIXS intensity attributed to charge transfer (energy loss $E>4.2$ eV RIXS intensity) is plotted in blue. Circles indicate (magenta) the first resonance state energy, and (red) the $M_{2,3}$ charge transfer threshold as defined in the text. Panels (d-f) show the same spectra for artificially doubled Slater-Condon values ($SC\times2$). (g) The energy difference between the first state and the $M_{2,3}$ threshold energy is plotted as a function of the core hole potential, and compared with (dashed line) a slope of -1.}
\label{fig:MedgeTH}
\end{figure*} 

Simulated RIXS spectra are obtained using the Kramers-Heisenberg equation (see Fig. \ref{fig:RIXSimages}), with exactly diagonalized eigenstates of the model Hamiltonian (as in Ref. \cite{KHmethod}). The VUV RIXS measurement in Fig. \ref{fig:cartoons}(a, right) was performed at the beamline 4.0.3 (MERLIN) RIXS endstation (MERIXS) at the Advanced Light Source (ALS), Lawrence Berkeley National Laboratory. The data were recorded by a VLS based X-ray emission spectrograph equipped with a commercially available CCD detector \cite{YiDeDetector,YiDeSRN}. A large single crystal of NiO was measured near room temperature at a pressure of 3$\times$10$^{-10}$ Torr. The $\pi$-polarized photon beam had a grazing 25$^o$ angle of incidence to the cleaved [100] sample face, and scattered photons were measured with a 90$^o$ included scattering angle. The RIXS simulations of charge transfer intensity were are also performed in this scattering geometry.

\section{Results and discussion}

The degree to which RIXS couples to charge transfer states reflects how the core hole resonance states deviate from the ground state charge density distribution. Figure \ref{fig:shakeVsU} evaluates how the total charge transfer intensity in RIXS depends on core hole potential and Slater-Condon interaction strengths, when charge transfer excitations are summed over the M- and L-edge resonances. The RIXS spectrum of NiO is particularly amenable to this study, because non-charge-transfer multiplet features occur entirely below $\lesssim$4 eV, within the charge transfer gap (see Fig. \ref{fig:RIXSimages}). As the core hole potential increases, charge transfer intensity increases from $\lesssim$20$\%$ of the incident-energy-averaged spectrum at $U=0$ to $\gtrsim$40$\%$ of the RIXS spectrum at $U=3$ eV. Doubling the higher order Slater-Condon interaction terms yields an effect similar to increasing $U$ by 1 eV and 0.5 eV at the M- and L-edges, respectively.

The fractional intensity of RIXS charge transfer excitations in these parameter ranges is broken down as a function of incident energy in Fig. \ref{fig:LedgeTH}-\ref{fig:MedgeTH}. Figure \ref{fig:LedgeTH}(a-f) shows that at the soft X-ray L-edge, ligand hybridization results in a weak charge transfer tail at the high energy edge of XAS. A characteristic 3-peak pattern in simulated charge transfer features can be observed due to the small number of discrete SAIM ligand band states (N=3), and reveals the degree to which ligand DOS inhomogeneity influences the line shape of these features. The onset of this high energy charge transfer tail is identified from points of maximum slope in the RIXS fractional charge transfer intensity at the $L_3$ and $L_2$ edges, and marked with red and black dots, respectively.

As expected from the trend in Fig. \ref{fig:shakeVsU}, the the XAS charge transfer tail becomes significantly more prominent as the core hole potential $U$ increases. Figure \ref{fig:LedgeTH}(b) represents a typical L-edge parameter set, for which the charge transfer tail is quite weak. The degree to which resonance states in this tail couple to RIXS charge transfer excitations also increases steadily with $U$, and is in all cases well above the value for states beneath the threshold ($I_{CT}\sim10-20\%$). The charge transfer threshold energy plotted in Fig. \ref{fig:LedgeTH}(g) is defined by subtracting the energy of the first dipole-allowed resonance state (magenta dot) from the charge transfer onset, and roughly follows an $m=-1$ slope. Doubling the higher order Slater-Condon interaction terms significantly enhances charge transfer in RIXS, but also causes the charge transfer threshold energy to grow by $\sim1$ eV. Increasing this energy scale thus has the effect of pushing higher energy states towards (or across) the charge transfer threshold, while pulling leading edge states further below it.

The core hole potential and Slater-Condon interaction parameter dependence of scattering at the VUV M-edge is shown in Fig. \ref{fig:MedgeTH}, and exhibits similar trends to the L-edge calculations in Fig. \ref{fig:LedgeTH}. Only one charge transfer threshold is observed, as core level spin orbit coupling is not strong enough to differentiate $M_3$ and $M_2$ edges. Larger Slater Condon interaction terms at the M-edge significantly increase the charge transfer threshold energy, and charge transfer modes are not very significant in the spectrum with realistic parameters (Fig. \ref{fig:MedgeTH}(a)). It is also of interest that the fractional intensity of charge transfer RIXS features within the high energy tail is consistently smaller at the M-edge than at the L-edge. Combined with the smaller $U$ values in the VUV, this suggests that the M-edge resonance states deviate less from a monovalent effective orbital picture (i.e. scattering models that neglect charge transfer will be more accurate in the VUV).

Sufficiently strong interactions will drive the higher energy multiplet peaks across the threshold, as is seen in Fig. \ref{fig:MedgeTH}(f). This scenario (driven by Slater-Condon interactions) is particularly common for VUV scattering from mid-transition metals and f-electron systems, and results in a much shorter ``autoionization" lifetime for the higher energy peaks as discussed in Ref. \cite{HaverkortAutoIon}. The ``autoionization lifetime" is the characteristic time required for charge transfer to occur in the presence of a core hole. In referring to this as a lifetime, we implicitly assume that an electron or hole delocalizes rapidly after leaving the scattering site, making it impossible for an autoionized core hole state to decay into low energy nominal-valence final states. However, other effects also contribute strongly to the phenomenon of higher energy features having broader line shapes in VUV spectroscopies \cite{KotaniIdea,KotaniIdea2,WrayNiO,WrayFrontiers,WrayRIXSinterference}. It should also be noted that though the model applied in this manuscript includes a mechanism for autoionization, it does not account for the full range of shake-up phenomenology that can be found in real materials (for examples, see Ref. \cite{WrayCoO,AmentRIXSReview}).



These results highlight the significant differences between 2-particle single atom density-density interactions described by a core hole potential and by Slater-Condon parameters as mechanisms for creating charge transfer shake-up excitations in RIXS. For the simulated scenarios, these interactions have very different effects on the charge transfer threshold, and VUV resonance generally preserves more intensity in low energy multiplet RIXS features. These simulations may also be significant for understanding VUV RIXS on actinides, which is a relatively new experimental frontier. It has been suggested that VUV resonance is necessary to measure clean multiplet-derived spectra on actinide compounds \cite{deGrootInvisible,WrayURS}, and mixed valence is critical to understanding the Kondo lattice and quasi-itinerant nature behavior common to these materials. Our analysis also has generic importance for high energy resolution RIXS studies enabled in the VUV.

\textbf{Acknowledgements:} The Advanced Light Source is supported by the Director, Office of Science, Office of Basic Energy Sciences, of the U.S. Department of Energy under Contract No. DE-AC02-05CH11231.



\end{document}